\documentclass[10pt]{article}

\usepackage{amsmath}
\usepackage{amssymb}

\usepackage{graphicx}

\usepackage{cite}

\usepackage{color}

\usepackage[labelfont=bf,labelsep=period,justification=raggedright]{caption}

\makeatletter
\renewcommand{\@biblabel}[1]{\quad#1.}
\makeatother

\date{}

\begin{document}

\begin{flushleft}
{\Large
\textbf{Evolution of scaling emergence in large-scale spatial epidemic
spreading}
}
\\
Lin Wang, Xiang Li$\ast$, Yi-Qing Zhang, Yan Zhang, Kan Zhang
\\
\bf Adaptive Networks and Control Lab, Department of Electronic Engineering,
Fudan University, Shanghai 200433, PR China
\\
$\ast$ Corresponding author: lix@fudan.edu.cn
\end{flushleft}

\section*{Abstract}

\textbf{Background:}
Zipf's law and Heaps' law are two representatives of the scaling concepts,
which play a significant role in the study of complexity science. The
coexistence of the Zipf's law and the Heaps' law motivates different
understandings on the dependence between these two scalings, which is still
hardly been clarified.

\textbf{Methodology/Principal Findings:}
In this article, we observe an evolution process of the scalings: the Zipf's law and the
Heaps' law are naturally shaped to coexist at the initial time, while the crossover comes
with the emergence of their inconsistency at the larger time before reaching a stable state,
where the Heaps' law still exists with the disappearance of strict Zipf's law. Such findings
are illustrated with a scenario of large-scale spatial epidemic spreading, and the empirical
results of pandemic disease support a universal analysis of the relation between the two laws
regardless of the biological details of disease. Employing the United States(U.S.) domestic
air transportation and demographic data to construct a metapopulation model for simulating the
pandemic spread at the U.S. country level, we uncover that the broad heterogeneity of the
infrastructure plays a key role in the evolution of scaling emergence.

\textbf{Conclusions/Significance:}
The analyses of large-scale spatial epidemic spreading help understand the
temporal evolution of scalings, indicating the coexistence of the Zipf's
law and the Heaps' law depends on the collective dynamics of epidemic
processes, and the heterogeneity of epidemic spread indicates the significance
of performing targeted containment strategies at the early time of a pandemic
disease.


\section*{Introduction}

Scaling concepts play a significant role in the field of complexity
science, where a considerable amount of efforts is devoted to understand these
universal properties underlying multifarious systems[1-4]. Two representatives
of scaling emergence are the Zipf's law and the Heaps' law. G.K. Zipf, sixty
years ago, found a power law distribution for the occurrence frequencies of
words within different written texts, when they were plotted in a descending order
against their rank[5]. This frequency-rank relation also corresponds to a power law
probability distribution of the word frequencies[32]. The Zipf's law is found to
hold empirically for a great deal of complex systems, e.g., natural and artificial
languages[5-9], city sizes[10,11], firm sizes[12], stock market index[13,14], gene
expression[15,16], chess opening[17], arts[18], paper citations[19], family names[20],
and personal donations[21]. Many mechanisms are proposed to trace the origin of the
Zipf's law[22-24].

Heaps' law is another important empirical principle describing the sublinear growth of
the number of unique elements, when the system size keeps on enlarging[25].
Recently, particular attention is paid to the coexistence of the Zipf's law and the Heaps' law,
which is reported for the corpus of web texts[26], keywords in scientific publication[27],
collaborative tagging in web applications[28,29], chemoinformatics[30], and more close
to the interest in this article, global pandemic spread[31], and etc.

In [33,34], an improved version of the classical Simon model[35] was put forward to investigate
the emergence of the Zipf's law, which is deemed to be a result from the existence of the Heaps'
law. However, [26,32] concluded that the Zipf's law leads to the Heaps' law. In fact, the
interdependence of these two laws has hardly been clarified. This embarrassment comes from the
fact that the empirical/simulated evidence employed to show the emergence of Zipf's law mainly
deals with static and finalized speicmens/results, while the Heaps' law actually describes the
evolving characteristics.

In this article, we investigate the relation between these scaling laws from the
perspective of coevolution between the scaling properties and the epidemic spread.
We take the scenarios of large-scale spatial epidemic spreading for example, since
the empirical data contain sufficient spatiotemporal information making it possible
to visualize the evolution of the scalings, which allows us to analyze the inherent
mechanisms of their formation. The Zipf's law and the Heaps' law of the laboratory
confirmed cases are naturally shaped to coexist during the early epidemic spread at
both the global and the U.S. levels, while the crossover comes with the emergence of
their inconsistency as the epidemic keeps on prevailing, where the Heaps' law still
exists with the disappearance of strict Zipf's law. With the U.S. domestic air
transportation and demographic data, we construct a fine-grained metapopulation model
to explore the relation between the two scalings, and recognize that the broad
heterogeneity of the infrastructure plays a key role in their temporal evolution,
regardless of the biological details of diseases.

\section*{Results}

\subsection*{Empirical and Analytical Results}

\begin{figure}[!ht]
\begin{center}
\includegraphics[width=6in]{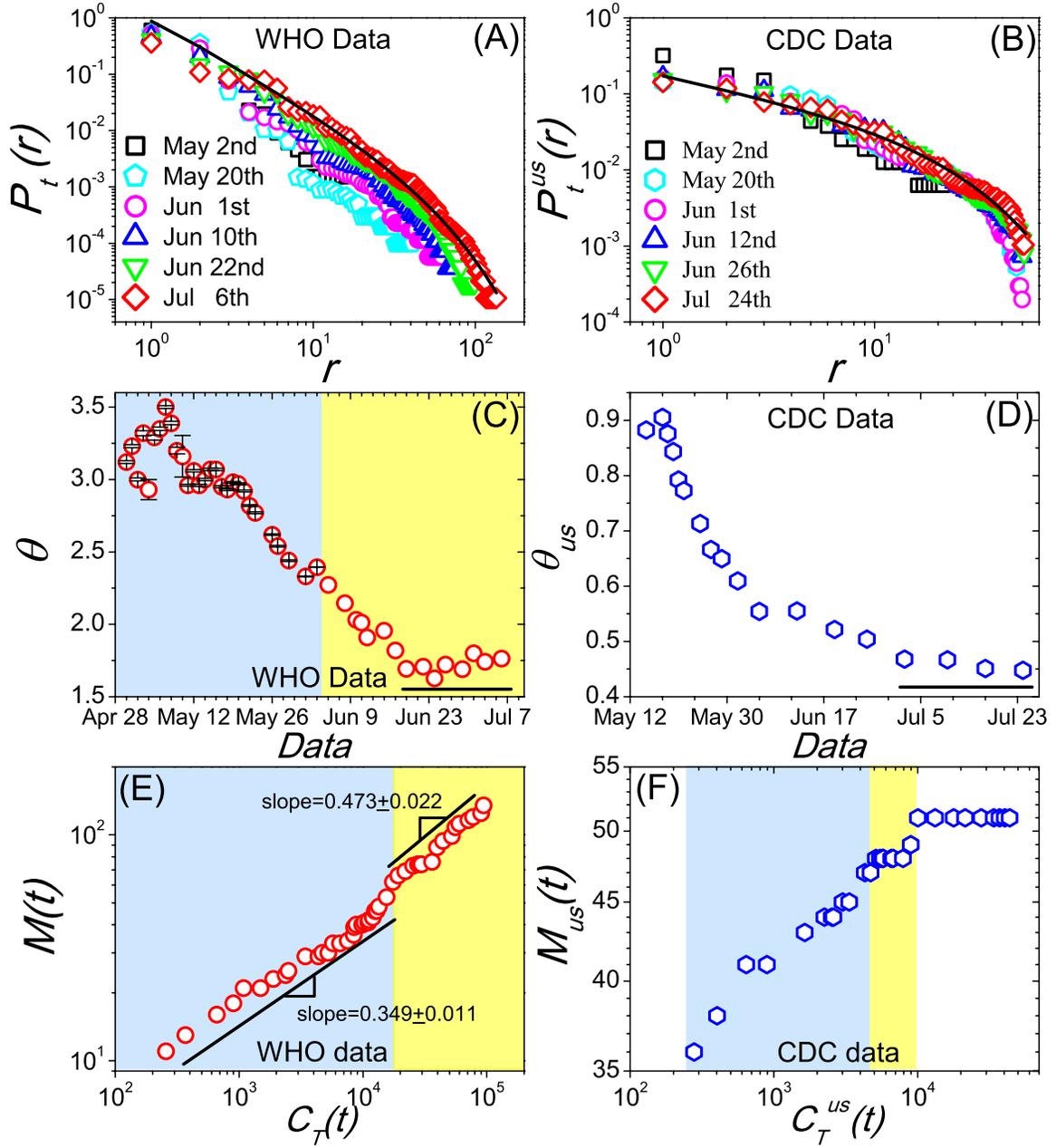}
\end{center}
\caption{\bf{The empirical results of A(H1N1).}
(A) The Zipf-plots of the normalized probability-rank distributions $P_{t}(r)$
of the cumulated confirmed number of every infected country at several given date sampled
about every two weeks, data provided by the WHO.
(B) The Zipf-plots of $P^{us}_{t}(r)$ at several given data sampled about every
two weeks, data provided by the CDC.
(C) Temporal evolution of the estimated exponent $\theta$ of the normalized
distribution $P_{t}(r)$.
(D) Temporal evolution of the estimated exponent $\theta_{us}$ of the normalized
distribution $P^{us}_{t}(r)$ of the period after May 15th.
(E) The sublinear relation between the number of infected countries $M(t)$
and the cumulative number of global confirmed cases $C_{T}(t)$, data collected by the WHO.
(F) The sublinear relation between the number of infected states $M_{us}(t)$
and the cumulative number of national confirmed cases $C^{us}_{T}(t)$, data collected by the CDC.
The shaded areas in the figures (C,E,F) corresponds to their different evolution
stages, respectively.}
\label{fig.1}
\end{figure}

With the empirical data of the laboratory confirmed cases of the A(H1N1) provided by
the World Health Organization(WHO)(see the data description in \emph{Materials and Methods}),
we first study the probability-rank distribution(\emph{PRD}) of the cumulative confirmed
number(\emph{CCN}) of every infected country at several given dates sampled about every two
weeks. $C_{j}(t)$ denotes the \emph{CCN} in a given country $j$ at time $t$. Since $C_{j}(t)$
grows with time, the distributions at different dates are normalized by the global \emph{CCN}, $C_{T}(t)=\sum_{j}C_{j}(t)$, for comparison. Fig.\ref{fig.1}(A) shows the Zipf-plots of the
\emph{PRD} $P_{t}(r)$ of the infected countries' confirmed cases by arranging every
$C_{j}(t)/C_{T}(t)>0$ in a descending order for each specimen. The maximal rank $r_{t, max}$(on
x-axis) for each specimen denotes the total number of infected countries at a given date, and
grows as the epidemic spreading.

At the early stage(the period between April 30th and June 1st, 2009), $P_{t}(r)$
shows a power law pattern $P_{t}(r)\sim r^{-\theta}$, which indicates the emergence of the
Zipf's law. We estimate the power law exponent $\theta$ for each specimen of this stage by
the maximum likelihood method[22,37], and report its temporal evolution in the left part of
Fig.\ref{fig.1}(C). About sixty countries were affected by the A(H1N1) on June 1st, and most
of them are countries with large population and/or economic power, e.g., U.S., Mexico, Canada,
Japan, Australia, China. After June 1st, the disease swept much more countries in a short time,
and the WHO announcement on June 11th[38] raised the pandemic level to its highest phase,
phase 6(see \emph{Text S1}), which implied that the global pandemic flu was
occurring. At this stage(after June 1st, 2009), $P_{t}(r)$ gradually displays a power law
distribution with an exponential cutoff $P_{t}(r)\sim r^{-\theta}exp(-r/r_{c})$, where $r_{c}$
is the parameter controlling the cutoff effect(see \emph{Text S1}), and the
exponent $\theta$ gradually reduces to around 1.7, as shown in Fig.\ref{fig.1}(C). Surprisingly,
$P_{t}(r)$ at different dates eventually reaches a stable distribution as time evolves(see those
curves since June in Fig.\ref{fig.1}(A)). Indeed, after June 19th, $\theta$ seems to reach a
stable value with mild fluctuations, as shown in Fig.\ref{fig.1}(C). The characteristics of the
temporal evolution of the parameter $r_{c}$ is similar to $\theta$, thus we mainly present the
empirical results of the exponent $\theta$ in the main text and hold the results of $r_{c}$ in
Figure S1. In the following, we analyze the evolution of the normalized distribution
$P_{t}(r)$ by the contact process of an epidemic transmission, regardless of the biological
details of diseases.

Straightforwardly, according to the mass action principle in the mathematical
epidemiology[39,40](see \emph{Text S1}), which is widely applied in studying
the epidemic spreading process on a network[41-56], we consider the SIR epidemic scheme here,
\begin{eqnarray}
(\mathcal {D}^{[S]}_{j},\mathcal {D}^{[I]}_{j},\mathcal
{D}^{[R]}_{j})\rightarrow\left\{\begin{array}{l@{} l}
\displaystyle(\mathcal {D}^{[S]}_{j}-1,\mathcal {D}^{[I]}_{j}+1,
\mathcal {D}^{[R]}_{j}),&\ with\ rate\  \beta \mathcal {D}^{[S]}_{j}\mathcal
{D}^{[I]}_{j}/N_{j},\\[0.1cm]
\displaystyle(\mathcal {D}^{[S]}_{j},\mathcal {D}^{[I]}_{j}-1,
\mathcal {D}^{[R]}_{j}+1),&\ with\ rate\  \mu \mathcal
{D}^{[I]}_{j},\\
\end{array}\right.\label{eq.1}
\end{eqnarray}
where $\mathcal {D}_{j}^{[\varphi]}$ denotes the number of individuals in compartment
$[\varphi]$(susceptible(S), infectious(I) or permanently recovered(R)) in a given
country $j$, $\beta$ denotes the disease transmission rate, and infectious individuals
recover with a probability $\mu$. The population in a given country $j$ at time $t$ is $N_{j}(t)=\sum_{\varphi}\mathcal{D}^{[\varphi]}_{j}(t)$, where $t=0$ means the time when
initially confirmed cases in the entire system are reported. At the early stage of a
pandemic outbreak, the new introductions of infectious individuals dominate the onset of
outbreak in unaffected countries. However, after the disease already lands in these
countries, the ongoing indigenous transmission gradually exceeds the influence of the
new introductions, and becomes the mainstream of disseminators[57,58]. According to
Eq.(\ref{eq.1}), in a given infected country $j$, there are
\begin{eqnarray}
\mathcal {D}^{[I_{new}]}_{j}(t+1)=\beta\mathcal {D}^{[S]}_{j}(t)\mathcal
{D}^{[I]}_{j}(t)/N_{j}(t)
\end{eqnarray}
new infected individuals on average at $t+1$ days, and the
average number of illness at $t+1$ days is
\begin{eqnarray}
\mathcal {D}^{[I]}_{j}(t+1)=(1-\mu+\beta\mathcal {D}^{[S]}_{j}(t)/N_{j}(t))\mathcal
{D}^{[I]}_{j}(t).
\end{eqnarray}
Defining $\chi(t)=-\mu+\beta\mathcal {D}^ {[S]}_{j}(t)/N_{j}(t)$
and $\mathcal {Y}(t)=\mathcal {D}^ {[S]}_{j}(t)/N_{j}(t)$, we have
\begin{eqnarray}
\mathcal {D}^{[I]}_{j}(t+1)=\prod^{t}\limits_{t'=t_{1}}[1+\chi(t')]\mathcal
{D}^{[I]}_{j}(t_{1}) \ \ and\ \  \mathcal {D}^{[I_{new}]}_{j}(t+1)=\beta\mathcal
{Y}(t)\prod\limits^{t-1}_{t'=t_{1}}[1+\chi(t')]\mathcal {D}^{[I]}_{j}(t_{1}),
\end{eqnarray}
where $\mathcal {D}^{[I]}_{j}(t_{1})$ denotes the number of initially confirmed or
introduced cases in country $j$, and is always a small positive integer.
The \emph{CCN} of country $j$ at $t+1$ days is
$C_{j}(t+1)=C_{j}(t)+\mathcal {D}^{[I_{new}]}_{j}(t+1)$. When $t$ is large enough, we have
\begin{eqnarray}
C_{j}(t+1)/C_{j}(t)=1+\beta\mathcal {Y}(t)\prod^{t-1}\limits_{t'=t_{1}}[1+\chi(t')]\mathcal
{D}^{[I]}_{j}(t_{1})/C_{j}(t).
\end{eqnarray}

Before the disease dies out in country $j$, $C_{j}(t)$ keeps increasing from the onset of
outbreak[59]. When $t$ is large enough, it is obviously $C_{j}(t)\gg 0$, $0\leqslant\mathcal
{Y}\ll 1, -\mu\leqslant\mathcal {X}(t')\ll \beta-\mu$, thus
$\prod^{t-1}\limits_{t'=t_{1}}[1+\chi(t')]$ is definitely larger than $0$ and can hardly be
infinity. $\mathcal {D}_{j}^{[I]}(t_{1})$ is a small positive integer, thus $\mathcal {D}_{j}^{[I]}(t_{1})/C_{j}(t)\sim 0$ when $t$ is large enough. We therefore
have $C_{j}(t+1)/C_{j}(t)\sim 1, j\leqslant M(t+1)$ for large $t$, where $M(t+1)$ is the total
number of infected countries after $t+1$ days of spreading. Thus the normalized probability
$P_{t+1}(r(j))$ at $t+1$ day is:
\begin{eqnarray}
P_{t+1}(r(j))=\frac{C_{j}(t+1)}{C_{T}(t+1)}=\frac{C_{j}(t)}
{\sum_{j}C_{j}(t)}=P_{t}(r(j)), j\leqslant M(t+1),\ with\ large\ t,\label{eq.6}
\end{eqnarray}
where $r(j)$ is the rank of the \emph{CCN} of country $j$ in the descending order of the
\emph{CCN} list of all infected countries. Eq.(\ref{eq.6}) indicates that each probability
$P_{t}(r(j))$ is invariant for large $t$, thus the normalized distribution $P_{t}(r)$
becomes stable when $t$ is large enough. The intrinsic reasons for the emergence of these
scaling properties are discussed in \emph{Modeling and Simulation Results}.

Since the normalized \emph{PRD} $P_{t}(r)$ displays the Zipf's law pattern $P_{t}(r)\sim
r^{-\theta}$ at the early stage of the epidemic, the \emph{CCN} of the country ranked $r$ is
$C_{r}(t)\sim C_{T}(t)\cdot r^{-\theta}$ at this stage. Considering the \emph{CCN} of the
countries with ranks between $r$ and $r+\delta r$, where $\delta r$ is any infinitesimal value,
we have $\delta C_{r}(t)\sim-\theta r^{-\theta-1}C_{T}(t)\delta r$. Supposing $\delta r\sim
P_{C_{r}}(t)\delta C_{r}(t)$ with $P_{C_{r}}$ denoting the probability density function, we have
\begin{eqnarray}
P_{C_{r}}(t)\sim-\theta^{-1}r^{\theta+1}C_{T}^{-1}(t).
\end{eqnarray}
Thus
\begin{eqnarray}
P_{C_{r}}(t)=\mathcal {A}(1-\phi)C_{T}^{\phi-1}(t)C^{-\phi}_{r}(t),
\end{eqnarray}
where $\phi=1+\theta^{-1}$, $\mathcal {A}$ is a constant. According to the normalization condition $\int_{C^{min}(t)}^{C^{max}(t)}P_{C_{r}}(t)dC_{r}(t)=1$, where $C^{max}(t)(C^{min}(t))$ is the
\emph{CCN} of the country with the maximal(minimal) value at a give time $t$, we have $\mathcal {A}=-C_{T}^{1-\phi}(t)C^{min}(t)^{\phi-1}$ because $\phi=1+\theta^{-1}>1$ and
$C^{max}(t)\gg 0$. Then
\begin{eqnarray}
P_{C_{r}}(t)=(\phi-1)C^{min}(t)^{\phi-1}C_{r}^{-\phi}(t).
\end{eqnarray}
At a given date, $r$ can be regarded as the number of countries with the amount of
cumulated confirmed cases which is no less than $C_{r}(t)$, then
\begin{eqnarray}
r=\int_{C_{r}(t)}^{C^{max}(t)}M(t)P(C_{r}'(t))dC_{r}'(t).
\end{eqnarray}
Recalling $r\sim (C_{T}(t)/C_{r}(t))^{\frac{1}{\theta}}$,  we have
\begin{eqnarray}
M(t)\sim(\frac{C_{T}(t)}{C^{min}(t)})^{\eta},\label{eq.3}
\end{eqnarray}
where $\eta=1/\theta$. At the early stage corresponding to the period between April 30th and June
1st, $C^{min}(t)$ is one according to the WHO data. Therefore, we have
\begin{eqnarray}
M(t)\sim C^{\eta}_{T}(t),\eta=1/\theta,\label{eq.4}
\end{eqnarray}
which indicates that the Heap's law[25,26,31,32] can be observed in this case. The
empirical evidence for the emergence of the Heap's law at this stage is shown in the
middle part of Fig.\ref{fig.1}(E). The Heaps' exponent $\eta$ is obtained by the least
square method[31,32], and the relevance between $\theta$ and $\eta$ is reported in
Table \ref{tab.1}.

At the latter stage(the period after June 1st, 2009), the exponential tail of the
distribution $P_{t}(r)$ leads to a deviation from the strict Zipf's law. However, with a steeper
exponent $\eta\approx 0.473$, the Heaps' law still exists, as shown in the right part of
Fig.\ref{fig.1}(E). Though the two scaling laws are naturally shaped to coexist during the early
epidemic spreading, their inconsistency gradually emerges as the epidemic keeps on prevailing.
Indeed, in the \emph{Discussion} of [32], without empirical or analytical evidence, L\"{u} et al
have intuitively suspected that there may exist some unknown mechanisms only producing the Heaps'
law, and it is possible that a system displaying the Heaps' law does not obey the strict Zipf's
law. Here we not only verify this suspicion with the empirical results, but also explore the substaintial
mechanisms of the evolution process in \emph{Modeling and Simulation Results}, where we uncover the
important role of the broad heterogeneity of the infrastructure in the temporal evolution of scaling
emergence.

We also empirically study the evolution of scaling emergence of the epidemic
spreading at the countrywide level. Since the United States is one of the several earliest
and most seriously prevailed countries of the A(H1N1)[60], we mainly focus on the A(H1N1)
spreading in the United States. With the empirical data of the laboratory confirmed cases
of the A(H1N1) provided by the Centers for Disease Control and Prevention(CDC)(see the data
description in \emph{Materials and Methods}), in Fig.\ref{fig.1}(B) we report the \emph{PRD}
of the \emph{CCN} of infected states, $P^{us}_{t}(r)$, at several given dates sampled about
every two weeks. Our findings suggest a crossover in the temporal evolution of $P^{us}_{t}(r)$.
At the early stage(the period before May 15th), $P^{us}_{t}(r)$ shows a power law pattern $P^{us}_{t}(r)\sim r^{-\theta_{us}}$ with a much smaller exponent $\theta_{us}$ than that of
the WHO results. Washington D.C. and 46 states(excluding Alaska, Mississippi, West Virginia,
Wyoming) were affected by A(H1N1) on May 15th. After May 15th, $P^{us}_{t}(r)$ gradually
becomes a power law distribution with an exponential cutoff,
$P^{us}_{t}(r)\sim r^{-\theta_{us}}exp(-r/r^{us}_{c})$, which leads to a deviation from the
strict Zipf's law. In this case, the exponent $\theta_{us}$ gradually reduces and reaches a
stable value 0.45(see Fig.\ref{fig.1}(D)), which conforms to the fact that $P^{us}_{t}(r)$ of
different dates eventually reaches a stable distribution as time evolves. The temporal evolution
of the exponent $\theta_{us}$ of all data are shown in Figure S2. $r_{c}$ keeps the value around
14 after June 12th, 2009.

The relation between $M_{us}(t)$ and $C^{us}_{T}(t)$ is shown in
Fig.\ref{fig.1}(F). Though at first glance this figure provides us an
impression of the sublinear growth of the number of infected states $M_{us}(t)$
when the cumulative number of national total patients $C^{us}_{T}(t)$ increases,
we could not use the least square method here to estimate the Heaps' exponent
$\eta_{us}$ for several reasons: (i) the amount of data at each stage is quite
small; (ii) there are several periods that $M_{us}(t)$ keeps unchanged(May 6th
$\rightarrow$ May 7th, $M_{us}(t)=41$; May 12th $\rightarrow$ May 13th,
$M_{us}(t)=45$; May 18th $\rightarrow$ May 27th, $M_{us}(t)=48$); (iii) the
magnitude of $C^{us}_{T}(t)$ is much larger than that of $M_{us}(t)$; (iv) after
June 1st, 2009, Washington D.C. and all 50 states of the United States were
affected by the A(H1N1). Define $M^{max}$ the maximal number of the geographical
regions the epidemic spreads to. In the U.S. scenario, $M^{max}_{us}=51$. When
$M_{us}(t)$ reaches $M^{max}_{us}$ on June 1st, $P^{us}_{t}(r)$ evolves and
becomes stable after June 26th(see Fig.\ref{fig.1}(B,D)). In the \emph{Modeling
and Simulation Results}, we explore the relation between these
two scalings with a fine grained metapopulation model characterizing the spread
of the A(H1N1) at the U.S. level in detail.

Note that these scaling properties are not exceptive for the A(H1N1) transmission.
More supported exemplifications are reported in \emph{Figure S3}, e.g. the
cases of SARS, Avian Influenza(H5N1). It is worth remarking that the normalized distribution
$P_{t}(r)$ almost keeps the power law pattern during the whole spreading process of the global
SARS. This phenomenon might result from the intense containment strategies, e.g. patient isolation,
enforced quarantine, school closing, travel restriction, implemented by individuals or governments
confronting mortal plague.

\subsection*{Modeling and Simulation Results}

The above analyses, however, do not tell the whole story, because the
intrinsic reasons for the emergence of these scaling properties have not been
explained. Some additional clues from the perspective of Shannon entropy[61] of
a system might unlock the puzzle.

Nowadays, population explosion in the urban areas, massive
interconnectivity among different geographical regions, and huge volume of
human mobility are the factors accelerating the spread of infectious
disease[62,74]. At a large geographical scale, one main class
of models is the metapopulation model dividing the entire system into several
interconnected subpopulations[58,63-74,87,88]. Within each subpopulation,
the infectious dynamics is described by the compartment schemes, while the
spread from one subpopulation to another is due to the transportation and
mobility infrastructures, e.g., air transportation. Individuals in each
subpopulation exist in various discrete health compartments(status), i.e.
susceptible, latent, infectious, recovered, and etc., with
compartmental transitions by the contagion process or spontaneous transition,
and might travel to other subpopulations by vehicles, e.g., airplane, in a short
time. The metapopulation model can not only be employed to describe
the global pandemic spread when we regard each subpopulation as a given country,
but also be used to simulate the disease transmission within a country when each
subpopulation is regarded as a given geographical region in the country. Here we
mainly consider the spread of pandemic influenza at the U.S. country level for
threefold reasons: (i) the computational cost of simulating global pandemic spread
is too tremendous to implement on a single PC or Server[58,70,72,81,87]; (ii) the
IATA or OAG flight schedule data, which is widely used to obtain the global air
transportation network, do not provide the attendance and flight-connecting
information(see data description in \emph{Materials and Methods}); (iii) the United
States is one of the several earliest and most seriously prevailed countries[60].

\begin{figure}[!ht]
\begin{center}
\includegraphics[width=3.5in]{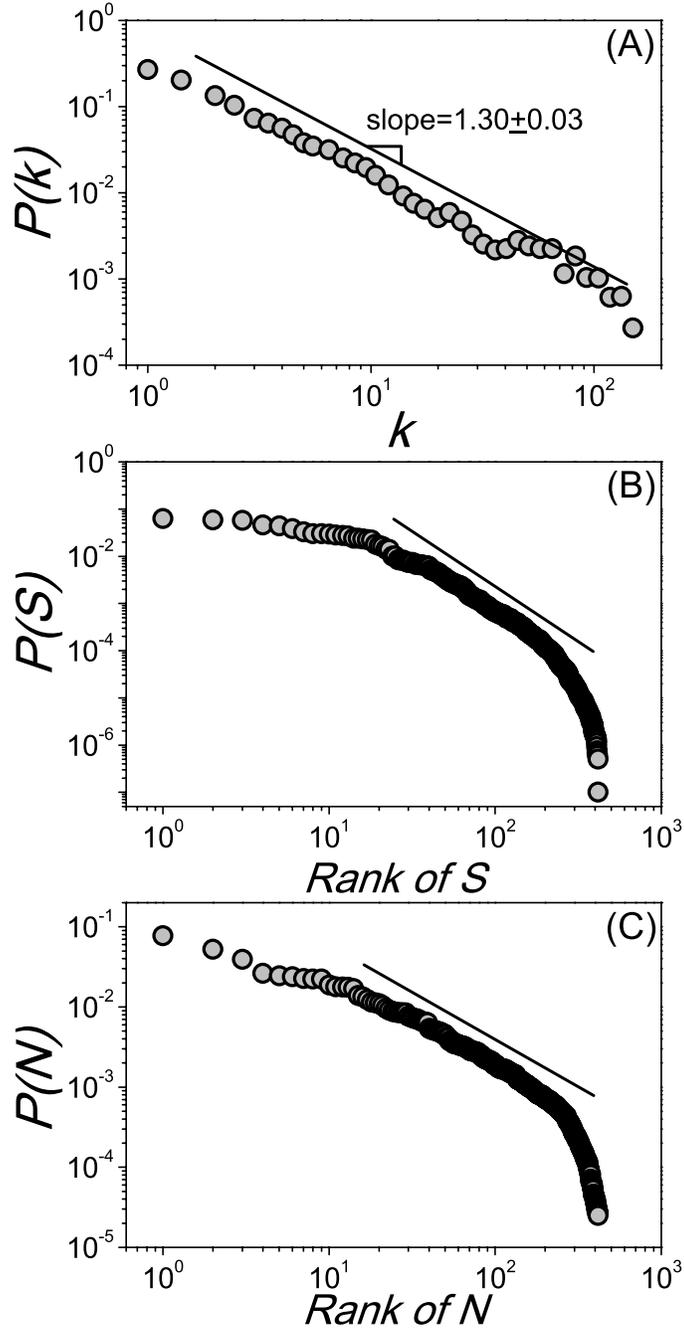}
\end{center}
\caption{\bf The heterogeneity of the USDAN's infrastructure.
(A) The degree distribution $P(k)$ follows a power law pattern on almost two
decades with an exponent 1.30$\pm$0.03. (B) shows that the probability-rank
distribution of the traffic outflux $S_{j}=\sum_{\ell\in\upsilon}\omega_{j\ell}$,
where $\upsilon$ denotes the set of neighbors belonging to the vertex $j$ and
the weight $\omega_{j\ell}$ of a connection between two vertices $(j,\ell)$ is
the number of passengers traveling a given route per day, is skewed and heterogeneously
distributed. (C) shows that the probability-rank distribution of populations
is skewed and heterogeneously distributed.}
\label{fig.2}
\end{figure}

We construct a metapopulation model at the U.S. level with the U.S. domestic air
transportation and demographic statistical data[75-78](detailed data description is
provided in \emph{Materials and Methods}, and a full specification of the simulation
model is reported in \emph{Text S1}). Define a subpopulation as a
Metropolitan/Micropolitan Statistical Areas(MSAs/$\mu$SAs)[75] connected by a transportation
network, in this article, the U.S. domestic airline network(USDAN). The USDAN is a weighted
graph comprising $V=406$ vertices(airports) and $E=6660$ weighted and directed edges denoting
flight courses. The weight of each edge is the daily amount of passengers on that
flight course. The infrastructure of the USDAN presents high levels of heterogeneity in
connectivity patterns, traffic capacities and population(see Fig.\ref{fig.2}). The disease
dynamics in a single subpopulation is modeled with the Susceptible-Latent-Infectious-Recovered(SLIR) compartmental scheme, where the abbreviation L denotes the latent compartment which experiences $\epsilon^{-1}$ days on average for an infected person(The SIR epidemic dynamics discussed at \emph{Empirical and Analytical Results} is an reasonable approximation, which actually simplifies the epidemic evolution to a Markov chain to help us study the issue, and the value of the reproductive number $R_{0}$ does not depend on $\epsilon$, we
therefore ignore the compartment L there).

The key parameters determining the spreading rate of infections are the reproductive number $R_{0}$ and the generation time $G_{t}$. $R_{0}$ is defined as the average amount of individuals an ill person infects during his or her infectious period $\mu^{-1}$ in a large fully susceptible population, and $G_{t}$ refers to the sum of the latent period $\epsilon^{-1}$ and the infectious period $\mu^{-1}$. In our metapopulation model, $R_{0}=\beta\cdot \mu^{-1}$. The initial conditions of the disease are defined as the onset of the outbreak in
San Diego-Carlsbad-San Marcos, CA MSA on April 17th, 2009, as reported by the CDC[79]. Assuming a short latent period value $\epsilon^{-1}=1.1$ days as indicated by the early estimates of the pandemic A(H1N1)[80], which is compatible with other recent studies[81,82], we primarily consider a baseline case with parameters: $G_{t}=3.6, \mu^{-1}=2.5$ days and $R_{0}=1.75$, which are higher than those obtained in the early findings of the pandemic A(H1N1)[80], but they are the median results in other subsequent analyses[81,83]. Fixing the latency period to $\epsilon^{-1}=1.1$ days, we also employ a more aggravated baseline scenario with parameters: $G_{t}=4.1, \mu^{-1}=3$ days and $R_{0}=2.3$, which are close to the upper bound results in[81,83-85].

In succession, we characterize the disease spreading pattern by information entropy,
which is customarily applied in information theory. To quantify the heterogeneity of
the epidemic spread at the U.S. level, we examine the prevalence at each time $t$, $i_{j}(t)={D}^{[I]}_{j}(t)/N_{j}(t)$, for all subpopulations, and introduce the
normalized vector $\vec{p}^{[i]}$ with components $p_{j}^{[i]}(t)=i_{j}(t)/\sum_{k}i_{k}(t)$.
Then we measure the level of heterogeneity of the disease prevalence by quantifying the
disorder encoded in $\vec{p}^{[i]}$ with the normalized entropy function
\begin{eqnarray}
H^{[i]}(t)=-\frac{1}{\log V}\sum_{j}p^{[i]}_{j}(t)\log
p^{[i]}_{j}(t),
\end{eqnarray}
which provides an estimation of the geographical heterogeneity of the disease spread at time $t$. If the
disease is uniformly influencing all subpopulations(e.g., all prevalences are equivalent), the entropy
reaches its maximum value $H^{[i]}=1$. On the other hand, starting from $H^{[i]}=0$, which is the most
localized and heterogeneous situation that just one subpopulation is initially affected by the disease, $H^{[i]}(t)$ increases as more subpopulations are influenced, thus decreasing the level of heterogeneity.

In order to better uncover the origin of the emergence of the scaling properties, we
compare the baseline results with those obtained on a null model \emph{UNI}. The
\emph{UNI} model is a homogeneous Erd\"{o}s-R\'{e}nyi random network with the same
number of vertices as that of the USDAN, and the generating regulation is described as
follows: for each pair of vertices $(i,j)$, an edge is independently generated with the
uniform probability $p_{e}=\langle k\rangle/V$, where $\langle k\rangle=16.40$ is the
average out-degree of the USDAN. Moreover, the weights of the edges and the populations
are uniformly equal to their average values in the USDAN, respectively. Therefore, the
\emph{UNI} model is completely absent from the heterogeneity of the airline topology, flux and population data.

\begin{figure}[!ht]
\begin{center}
\includegraphics[width=6in]{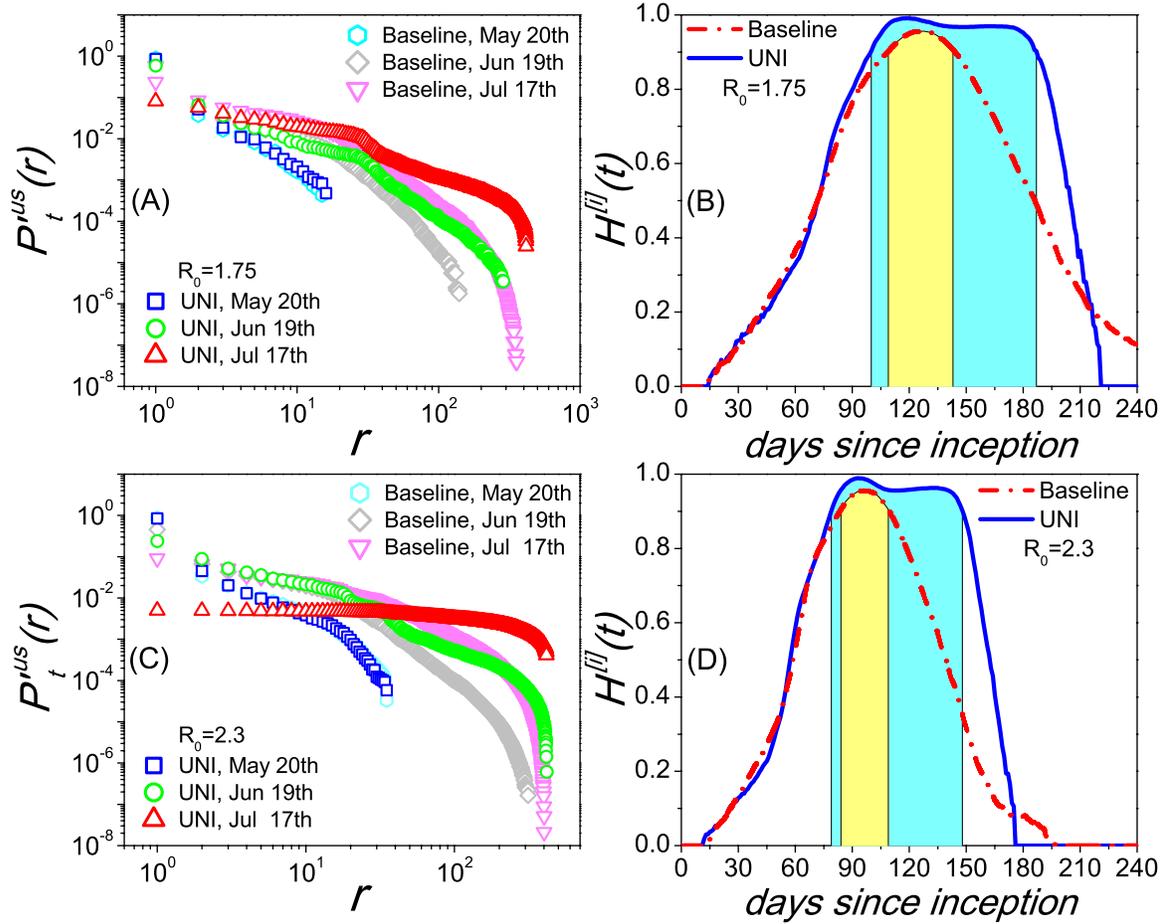}
\end{center}
\caption{\bf Comparisons of the scaling properties between the \emph{UNI}
scenarios and the baseline cases.
(A,C) present the comparison of the \emph{PRD} $P'^{us}_{t}(r)$
of the \emph{CCN} of every infected \emph{MSA/$\mu$SA} between the baselines and
the \emph{UNI} scenarios at several given date sampled about every 30 days when
$R_{0}=1.75, R_{0}=2.3$, respectively.
(B,D) present the comparison of the information entropy profiles between
the baselines and the \emph{UNI} results when $R_{0}=1.75, R_{0}=2.3$, respectively.
Each data in these figures are the median results over all runs that
led to an outbreak at the U.S. level in 100 random Monte Carlo realizations.}
\label{fig.3}
\end{figure}

Different evolving behaviors between the \emph{UNI} scenarios and the baselines(real
airline cases) provide a remarkable evidence for the direct dependence between the
scaling toproperties and the heterogeneous infrastructure. Fig.\ref{fig.3}(A,C) show
the comparison of the \emph{PRD} between the baseline results and the \emph{UNI}
outputs at several given dates sampled about every 30 days, where each specimen is the
median result over all runs that led to an outbreak at the U.S. level
in 100 random Monte Carlo realizations. In Fig.\ref{fig.3}(A), we consider the situation
of $R_{0}=1.75$, and do observe that the evolution of \emph{PRD} of the baseline case
experiences two stages: a power law at the initial time and an exponentially cutoff power
law at a larger time. However, the \emph{UNI} scenario shows a distinct pattern: as time
evolves, the middle part of the \emph{PRD} grows more quickly, and displays a peak which
obviously deviates scaling properties. Fig.\ref{fig.3}(C) reports the situation of
$R_{0}=2.3$. In this aggravated instance, the \emph{PRD} of the \emph{UNI} scenario actually
becomes rather homogeneous when $t$ is large enough(see the curve of July 17th of the
\emph{UNI} scenario in Fig.\ref{fig.3}(C)). Fig.\ref{fig.3}(B,D) present the comparison
of the information entropy profiles between the baseline results and the \emph{UNI}
outputs when $R_{0}=1.75, R_{0}=2.3$, respectively. The completely homogeneous network
\emph{UNI} shows a homogeneous evolution($H^{[i]}\approx1$) of the epidemic spread in
a long period(see the light cyan areas in Fig.\ref{fig.3}(B,D)), with sharp fallings at
both the beginning and the end of the outbreak. However, we observe distinct results in
the baselines, where $H^{[i]}$ is significantly smaller than 1 for most of the time,
and the long tails indicate a long lasting heterogeneity of the epidemic prevalence.
These analyses signal that the broad heterogeneity of infrastructure plays an essential
role in the emergence of scalings.

\begin{figure}[!ht]
\begin{center}
\includegraphics[width=5in]{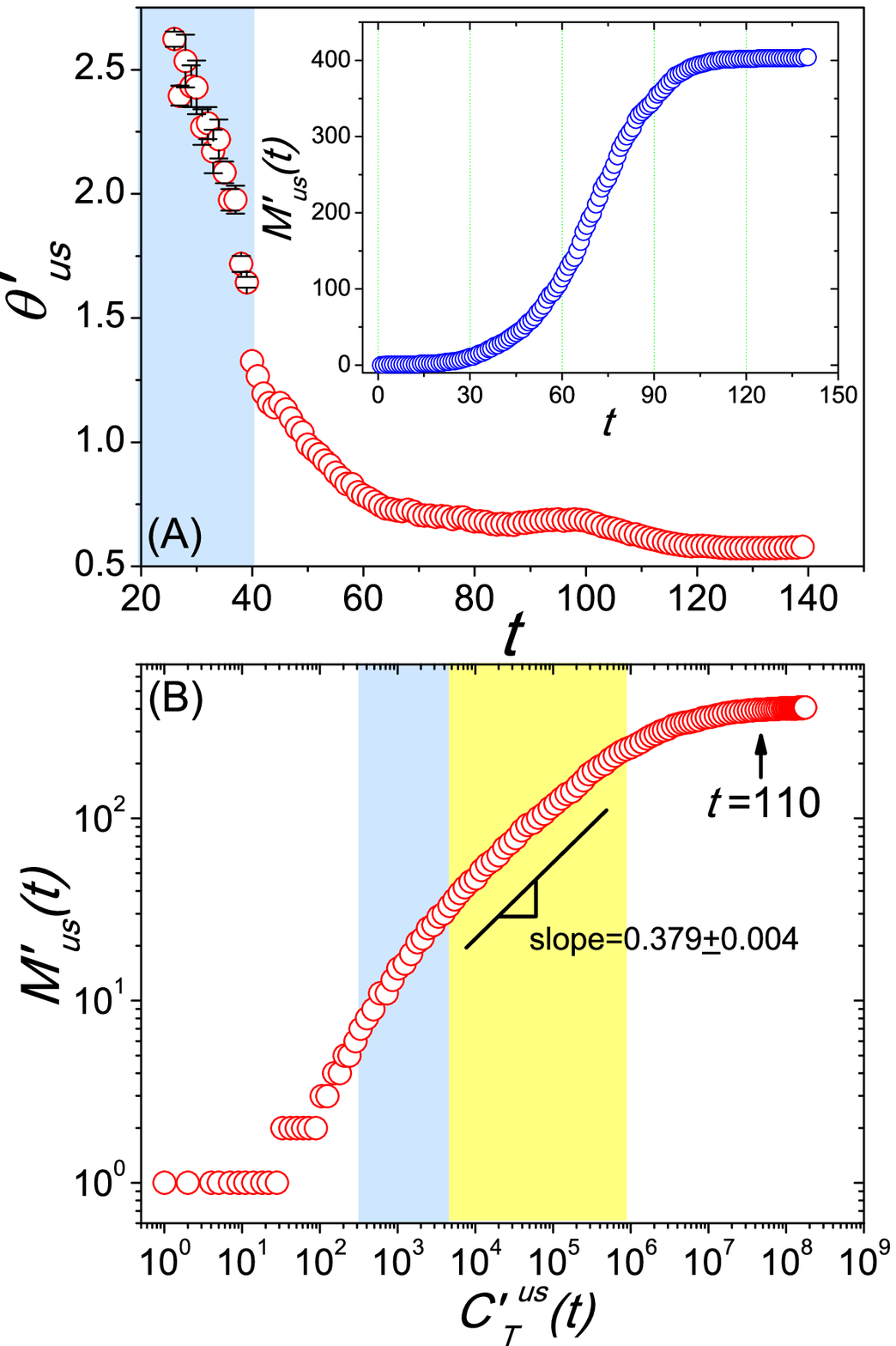}
\end{center}
\caption{\bf The statistical results of the scaling properties of our
metapopulation model.
(A) Temporal evolution of the estimated exponent $\theta'_{us}$ of the
normalized distribution $P'^{us}_{t}(r)$. The inset shows the growing of the number
of infected subpopulations $M'_{us}(t)$ with time $t$.
(B) The relation between the number of infected subpopulations $M'_{us}(t)$
and the national cumulative confirmed cases $C'^{us}_{T}(t)$.
The shaped areas in the figures corresponds to their different evolution
stages, respectively.
Each data in these figures are the median results over all runs that led
to an outbreak at the U.S. level in 100 random Monte Carlo realizations.}
\label{fig.4}
\end{figure}

We further explore the properties of the two scalings and their relation
with the baseline case of $R_{0}=1.75$ in detail. Since each independent simulation
generates a stochastic realization of the spreading process, we analyze the statistical
properties with 100 random Monte Carlo realizations, measure the normalized \emph{PRD}
of the \emph{CCN} of infected MSAs/$\mu$SAs for each realization that led to an outbreak
at the U.S. level, and report the median result of the \emph{PRD} $P'^{us}_{t}(r)$ of each
day. From $t=26$ to $t=39$, $P'^{us}_{t}(r)$ clearly shows a power law pattern
$P'^{us}_{t}(r)\sim r^{-\theta'_{us}}$, which implies the emergence of the Zipf's law(when
$t<26$, just several regions are affected by the disease). The exponent $\theta'_{us}$ at
each date is estimated by the maximum likelihood method[22,37], and the temporal evolution
of $\theta'_{us}$ is reported in the left part of Fig.\ref{fig.4}(A). When $t>39$,
$P'^{us}_{t}(r)$ gradually becomes an exponentially cutoff power law distribution
$P'^{us}_{t}(r)\sim r^{-\theta'_{us}}exp(-r/r^{us'}_{c})$, and the exponent $\theta'_{us}$
gradually reduces and reaches a stable value of 0.574 with neglectable fluctuations when
$t>126$(see Fig.\ref{fig.4}(A)). Here we do not show the error bar since the fitting error on the
exponent is far less($10^{-2}$) than the value of $\theta'_{us}$ by the average of 100 random
realizations. The inset of Fig.\ref{fig.4}(A) shows the increase of the number of infected regions
$M'_{us}(t)$ as time evolves. When $t>110$, more than 400 subpopulations reports the existence of
confirmed cases, thus $M'_{us}(t)$ tends to reach its saturation.

Fig.\ref{fig.4}(B) shows the relation between $M'_{us}(t)$ and $C'^{us}_{T}(t)$(the
national cumulative number of patients). Since $P'^{us}_{t}(r)$ displays a power law of
$P'^{us}_{t}(r)=b\cdot r^{\theta'_{us}}$ at the early stage of the period between $t=26$ and $t=39$,
it is reasonable to deduce the existence of the Heaps' law
\begin{eqnarray}
M'_{us}(t)=(C'^{us}_{T}(t)\cdot b)^{\eta'_{us}},\eta'_{us}=1/\theta'_{us},\label{eq.14}
\end{eqnarray}
according to the analyses in \emph{Empirical and Analytical Results}. In order to verify this
assumption, we estimate the exponent $\eta'_{us}$ using Eq.(\ref{eq.14}), and report the relevance
between $\theta'_{us}$ and $\eta'_{us}$ in Table 2(the amount of data in this period is not sufficient
to get a accurate estimation of the exponent $\eta'_{us}$ with the least square method). When $t>39$,
though $P'^{us}_{t}(r)$ gradually deviates the strict Zipf's law, the Heaps' law of the relation
between $M'_{us}(t)$ and $C'^{us}_{T}(t)$ still exists till $M'_{us}(t)$ tends to reach its
saturation(see the middle part in Fig.\ref{fig.4}(B)).

\section*{Discussion}

Zipf's law and Heaps' law are two representatives of the scaling concepts in
the study of complexity science. Recently, increasing evidence of the coexistence of the Zipf's
law and the Heaps' law motivates different understandings on the dependence between these two
scalings, which is still hardly been clarified. This embarrassment derives from the contradiction
that the empirical or simulated materials employed to show the emergence of Zipf's law are often
finalized and static specimens, while the Heaps' law actually describes the evolving characteristics.

In this article, we have identified the relation between the Zipf's law and the Heaps' law
from the perspective of coevolution between the scalings and large-scale spatial epidemic spreading.
We illustrate the temporal evolution of the scalings: the Zipf's law and the Heaps' law are naturally shaped to coexist at the early stage of the epidemic at both the global and the U.S. levels, while the crossover comes with the emergence of their inconsistency at a larger time before reaching a stable state, where the Heaps' law still exists with the disappearance of strict Zipf's law.

With the U.S. domestic air transportation and demographic data, we construct a
metapopulation model at the U.S. level. The simulation results predict main empirical findings.
Employing information entropy characterizing the epidemic spreading pattern,
we recognize that the broad heterogeneity of the infrastructure plays an essential role in the
evolution of scaling emergence. These findings are quite different from the previous conclusions
in the literature. For example, studying a phenomenologically self-adaptive complete network, Han
et al. claimed that scaling properties are dependent on the intensity of containment strategies
implemented to restrict the interregional travel[31]. In [36], Picoli Junior et al. considered a
simple stochastic model based on the multiplicative process[23], and suggested that seasonality
and weather conditions, i.e., temperature and relative humidity, also dominates the temporal
evolution of scalings because they affect the dynamics of influenza transmission. In this work,
without the help of any specific additional factor, we directly show that the evolution of scaling
emergence is mainly determined by the contact process underlying disease transmission on an
infrastructure with huge volume and heterogeneous structure of population flows among different
geographic regions. (The effects of the travel-related containment strategies implemented in real
world can be neglected, since the number of scheduled domestic and international passengers of
the U.S. air transportation only declined in 2009 by 5.3\% from 2008[86]. In fact, the travel
restrictions would not be able to significantly slow down the epidemic spread unless more
than 90\% of the flight volume is reduced[58,66,69,70,88].)

In summary, our study suggests that the analysis of large-scale spatial epidemic spread
as a promising new perspective to understand the temporal evolution of the scalings.
The unprecedented amount of information encoded in the empirical data of pandemic spreading
provides us a rich environment to unveil the intrinsic mechanisms of scaling emergence. The
heterogeneity of epidemic spread uncovered by the metapopulation model indicates the
significance of performing targeted containment strategies, e.g. vaccination of prior
groups, targeted antiviral prophylaxis, at the early time of a pandemic disease.

\section*{Materials and Methods}

\subsection*{Data Description}

In this article, in order to construct the U.S. domestic air transportation
network, we mainly utilize the ``\emph{Air Carrier Traffic and Capacity Data by On-Flight
Market report(December 2009)}" provided by the Bureau of Transportation Statistics(BTS)
database[76]. This report contains 12 months' data covering more than $96\%$
of the entire U.S. domestic air traffic in 2009, and provides the monthly number of
passengers, freight and/or mail transported between any two airports located within the
U.S. boundaries and territories, regardless of the number of stops between them. This
\emph{BTS} report provides a more accurate solution for studying aviation flows between
any two U.S. airports than other data sources(the attendance and the flight-connecting
information in the OAG flight schedule data are commonly unknown, while the datasets adopted
in [63,64,66,69] primarily consider the international passengers). In order to study the
epidemic spread in the Continental United States where we have a good probability to select
citizens living and moving in the mainland, we get rid of the airports as well as the
corresponding flight courses located in Hawaii, and all offshore U.S. territories and
possessions from the \emph{BTS} report.

In order to obtain the U.S. demographic data, we resort to the ``\emph{OMB Bulletin
N0. 10-02: Update of Statistical Area Definitions and Guidance on Their Uses}"[75]
provided by the United States Office of Management and Budget(OMB), and the
``\emph{Annual Estimates of the Population of Metropolitan and Micropolitan
Statistical Areas: April 1, 2000 to July 1, 2009}"[77] provided by the United States
Census Bureau(CB). OMB defines a Metropolitan Statistical Area(MSA)(Micropolitan
Statistical Area, $\mu$SA) as one or more adjacent counties or county equivalents that
have at least one urban core area of at least 50,000 population(10,000 population
but less than 50,000), plus adjacent territory that has a high degree of social and
economic integration with the core. For other regions with at least 5,000 population
but less than 10,000, we use the American FactFinder[78] provided by the CB to get the
demographic information. We do not consider sparsely populated areas with population
less than 5,000, because they are commonly remote islands, e.g. Block Island in Rhode
Island, Sand Point in Alaska.

Before constructing the metapopulation model, we take into account
the fact that there might be more than one airport in
some huge metropolitan areas. For instance, New York-Northern New
Jersey-Long Island(NY-NJ-PA MSA) has up to six airports(their IATA
codes: JFK, LGA, ISP, EWR, HPN, FRG), Los Angeles-Long Beach-Santa
Ana(CA MSA) has four airports(their IATA codes: LAX, LGB, SNA, BUR), and
Chicago-Joliet-Naperville(IL-IN-WI MSA) has two airports(their IATA codes:
MDW, ORD). Assuming a homogeneous mixing inside each subpopulation, we
need to assemble each group of airports serving the same MSA/$\mu$SA,
because the mixing within each given census areas is quite high and
cannot be characterized by fine-grained version of subpopulations for
every single airport. We searched for groups of airports located
close to each other and belonged to the same metropolitan areas, and
then manually aggregated the airports of the same group in a single
``super-hub".

The full list of updates of the pandemic A(H1N1) human cases of different
countries is available on the website of Global Alert and Response(GAR) of
World Health Organization(WHO)(WHO website. http://www.who.int/csr/disease/swineflu/updates/en/index.html. Accessed 2011 May 24).
It is worth remarking that WHO was no longer updating the number
of the cumulated confirmed cases for each country after July 6th, 2009, but changed
to report the number of confirmed cases on the WHO Region level(the
Member States of the World Health Organization(WHO) are grouped into six
regions, including WHO African Region(46 countries), WHO European Region(53
countries), WHO Eastern Mediterranean Region(21 countries), WHO Region of the
Americas(35 countries), WHO South-East Asia Region (11 countries), WHO Western
Pacific Region(27 countries).(WHO website.\\
http://www.who.int/about/regions/en/index.html. Accessed 2011 May 24).

The cumulative number of the laboratory confirmed human cases of A(H1N1) flu infection
of each U.S. state is available at the website of 2009 A(H1N1) Flu of the Centers for
Disease Control and Prevention(CDC)(CDC website. http://cdc.gov/h1n1flu/updates/. Accessed
2011 May 24), where the detailed data were started from April 23, 2009, to July 24, 2009.
After July 24, the CDC discontinued the reporting of individual confirmed cases of A(H1N1),
and began to report the total number of hospitalizations and deaths weekly.

The data of the human cases of global SARS and global Avian influenza(H5N1) are
available at the website of the Disease covered by GAR of WHO(WHO website. http://www.who.int/csr/disease/en/. Accessed 2011 May 24).




\section*{Acknowledgments}
We were grateful to the insightful comments of the editor
Alejandro Raul Hernandez Montoya and the two anonymous referees, and gratefully
acknowledged the helpful discussions with Changsong Zhou, Xiao-Pu Han, Zhi-Hai Rong,
Zhen Wang, Yang Yang. We also thank the Bureau of Transportation Statistics(BTS)
for providing us the U.S. domestic air traffic database.

\section*{References}

\ \ \ 1. Stanley HE (1999) Scaling, universality, and renormalization:
Three pillars of modern critical phenomena. Rev Mod Phys 71:
S358-S366.

2. Stanley HE, Amaral LAN, Gopikrishnan P, Ivanov PC, Keitt TH, et al
(2000) Scale invariance and universality: organizing principles in
complex systems. Physica A 281: 60-68.

3. Cardy J (1996) Scaling and Renormalization in Statistical Physics(Cambridge
University Press, New York).

4. Brown JH, West GB (2000) Scaling in Biology(Oxford University Press, USA).

5. Zipf GK (1949) Human Behaviour and the Principle of Least Effort:
An Introduction to Human Ecology(Addison-Wesley, Massachusetts).

6. Ferrer-i-Cancho R, Elvev\r{a}g B (2010) Random Texts Do Not Exhibit
the Real Zipf's Law-Like Rank Distribution. PLoS ONE 5: e9411.

7. Lieberman E, Michel JB, Jackson J, Tang T, Nowak MA (2007)
Quantifying the evolutionary dynamics of language. Nature 449: 713-716.

8. Kanter I, Kessler DA (1995) Markov Processes: Linguistics and
Zipf's Law. Phys Rev Lett 74, 4559-4562.

9. Maillart T, Sornette D, Spaeth S, von Krogh G (2008) Empirical Tests
of Zipf's Law Mechanism in Open Source Linux Distribution. Phys Rev
Lett 101, 218701.

10. Decker EH, Kerkhoff AJ, Moses ME (2007) Global Patterns of City Size
Distributions and Their Fundamental Drivers. PLoS ONE 2: e934.

11. Batty M (2006) Rank clocks. Nature 444: 592-596.

12. Axtell RL (2001) Zipf Distribution of U.S. Firm sizes. Science 293:
1818-1820.

13. Coronel-Brizio HF, Hern\'{a}ndez-Montoya AR (2005) On Fitting the
Pareto-Levy distribution to financial data: Selecting a suitable
fit's cut off parameter. Physica A 354: 437-449.

14. Coronel-Brizio HF, Hern\'{a}ndez-Montoya AR (2005) Asymptotic
behavior of the Daily Increment Distribution of the IPC, the Mexican
Stock Market Index. Revista Mexicana de F\'{i}sica 51: 27-31.

15. Ogasawara O, Okubo K (2009) On Theoretical Models of Gene Expression
Evolution with Random Genetic Drift and Natural Selection. PLoS ONE 4:
e7943.

16. Furusawa C, Kaneko K (2003) Zipf's Law in Gene Expression. Phys
Rev Lett 90: 088102.

17. Blasius B, T\"{o}njes R (2009) Zipf's Law in the Popularity
Distribution of Chess Openings. Phys Rev Lett 103: 218701.

18. Mart\'{i}nez-Mekler G, Mart\'{i}nez RA, del R\'{i}o MB, Mansilla R,
Miramontes P, et al. Universality of Rank-Ordering Distributions in the
Arts and Sciences. PloS ONE 4: e4791.

19. Redner S (1998) How popular is your paper? An empirical study of the
citation distribution. Eur Phys J B 4: 131-134.

20. Baek SK, Kiet HAT, Kim BJ (2007) Family name distributions: Master
equation approach. Phys Rev E 76: 046113.

21. Chen Q, Wang C, Wang Y (2009) Deformed Zipf's law in personal
donation. Europhys Lett 88: 38001.

22. Newman MEJ (2005) Power laws, Pareto distributions and Zipf's law.
Contemporary Physics 46: 323-351.

23. Sornette D (1997) Multiplicative processes and power laws.
Phys Rev E 57: 4811-4813.

24. Saichev A., Malevergne Y, Sornette D (2009) Theory of Zipf's
Law and Beyond, Lecture Notes in Economics and Mathematical
Systems(Springer).

25. Heaps HS (1978) Information Retrieval: Computational and Theoretical
Aspects(Academic Press, Orlando).

26. Serrano M\'{A}, Flammini A, Menczer F (2009) Modeling Statistical
Properties of Written Text. PLoS ONE 4: e5372.

27. Zhang ZK, L\"{u} L, Liu JG, Zhou T (2008) Empirical analysis on
a keyword-based semantic system. Eur Phys J B 66: 557-561.

28. Cattuto C, Barrat A, Baldassarri A, Schehr G, Loreto V (2009)
Collective dynamics of social annotation. Proc Natl Acad Sci 106:
10511-10515.

29. Cattuto C, Loreto V, Pietronero L (2007) Semiotic dynamics and
collaborative tagging. Proc Natl Acad Sci 104: 1461-1464.

30. Benz RW, Swamidass SJ, Baldi P (2008) Discovery of power-law in
chemical space. J Chem Inf Model 48: 1138-1151.

31. Han XP, Wang BH, Zhou CS, Zhou T, Zhu JF (2009) Scaling in the
Global Spreading Patterns of Pandemic Influenza A and the Role of Control:
Empirical Statistics and Modeling. eprint arXiv: 0912.1390.

32. L\"{u} L, Zhang ZK, Zhou T (2010) Zipf's Law Leads to Heaps' Law:
Analyzing Their Relation in Finite-Size Systems. PLoS ONE 5: e14139.

33. Montemurro MA, Zanette DH (2002) New perspectives on Zipf's law in
linguistics: from single texts to large corpora. Glottometrics 4: 86-98.

34. Zanette DH, Montemurro MA (2005) Dynamics of Text Generation with
Realistic Zipf's Distribution. J Quant Linguistics 12: 29-40.

35. Simon HA (1955) On a class of skew distribution functions. Biometrika
42: 425-440.

36. Picoli Junior Sd, Teixeira JJV, Ribeiro HV, Malacarne LC, Santos RPBd,
et al. (2011) Spreading Patterns of the Influenza A (H1N1) Pandemic. PLoS
ONE 6: e17823.

37. Clauset A, Shalizi CR, Newman MEJ (2009) Power-law distributions
in empirical data. SIAM Review 51, 661-703.

38. ``World now at the start of 2009 influenza pandemic", Statement to the
press by WHO Director-General Dr. Margaret Chan(June 11, 2009), World
Health Organization. WHO website.\\
http://www.who.int/mediacentre/news/statements/2009/h1n1\_{pandemic}\_{phase6}\_{20090611}/en/.
Accessed 2011 May 24.

39. Anderson RM, May RM (1991) Infectious Diseases of Humans: Dynamics
and Control(Oxford Unvi. Press, Oxford).

40. Hamer WH (1906) The Milroy Lectures On Epidemic disease in England ---
The evidence of variability and of presistency of type. The Lancet 167:
733-739.

41. Pastor-Satorras R, Vespignani A (2001) Epidemic Spreading in
Scale-Free Networks. Phys Rev Lett 86: 3200-3203.

42. Egu\'{i}luz VM, Klemm K (2002) Epidemic Threshold in Structured
Scale-Free Networks. Phys Rev Lett 89: 108701.

43. Barth\'{e}lemy M, Barrat A, Pastor-Satorras R, Vespignani A
(2004) Velocity and Hierarchical Spread of Epidemic Outbreaks in
Scale-Free Networks. Phys Rev Lett 92: 178701.

44. Gross T, D'Lima CJD, Blasius B (2006) Epidemic Dynamics on an
Adaptive Network. Phys Rev Lett 96: 208701.

45. Li X, Wang XF (2006) Controlling the spreading in small-world evolving
networks: stability, oscillation, and topology. IEEE T AUTOMAT CONTR 51:
534-540.

46. Zhou T, Liu JG, Bai WJ, Chen GR, Wang BH (2006) Behaviors of
susceptible-infected epidemics on scale-free networks with identical
infectivity. Phys Rev E 74: 056109.

47. Han XP (2007) Disease spreading with epidemic alert on small-world networks.
Phys Lett A 365: 1-5.

48. Yang R, Zhou T, Xie YB, Lai YC, Wang BH (2008) Optimal contact process
on complex networks. Phys Rev E 78: 066109.

49. Parshani R, Carmi S, Havlin S (2010) Epidemic Threshold for the
Susceptible-Infectious-Susceptible Model on Random Networks. Phys Rev
Lett 104: 258701.

50. Castellano C, Pastor-Satorras R (2010) Thresholds for Epidemic
Spreading in Networks. Phys Rev Lett 105: 218701.

51. Li X, Cao L, Cao GF (2010) Epidemic prevalence on random mobile
dynamical networks: Individual heterogeneity and correlation. Eur
Phys J B 75: 319-326.

52. Pulliam JR, Dushoff JG, Levin SA, Dobson AP (2007) Epidemic
Enhancement in Partially Immune Populations. PLoS ONE 2: e165.

53. Scoglio C, Schumm W, Schumm P, Easton T, Roy Chowdhury S, et al.
(2010) Efficient Mitigation Strategies for Epidemics in Rural Regions.
PLoS ONE 5: e11569.

54. Matrajt L, Longini IM Jr (2010) Optimizing Vaccine Allocation at
Different Points in Time during an Epidemic. PLoS ONE 5: e13767.

55. Iwami S, Suzuki T, Takeuchi Y (2009) Paradox of Vaccination: Is
Vaccination Really Effective against Avian Flu Epidemics? PLoS ONE 4:
e4915.

56. Bettencourt LMA, Ribeiro RM (2008) Real Time Bayesian Estimation
of the Epidemic Potential of Emerging Infectious Diseases. PLoS ONE 3:
e2185.

57. Longini IM Jr., Nizam A, Xu S, Ungchusak K, Hanshaoworakul W, et al
(2005) Containing Pandemic Influenza at the Source. Science 309: 1083-1087.

58. Bajardi P, Poletto C, Ramasco JJ, Tizzoni M, Colizza V, et al. (2011)
Human Mobility Networks, Travel Restrictions, and the Global Spread of 2009
H1N1 Pandemic. PLoS ONE 6: e16591.

59. Fraser C, Riley S, Anderson RM, Ferguson NM (2004) Factors that
make an infectious disease outbreak controllable. Proc Natl Acad Sci
USA 101: 6146-6151.

60. Situation updates---Pandemic (H1N1) 2009, World Health Organization.
WHO website.\\
http://www.who.int/csr/disease/swineflu/updates/en/index.html. Accessed
2011 May 24.

61. Shannon CE, Weaver W (1964) The Mathematical Theory of
Communication(The University of Illinois Press, Urbana).

62. Barab\'{a}si AL (2010) Bursts: The Hidden Pattern Behind Everything
We Do(Dutton Books, USA).

63. Rvachev LA, Longini IM Jr (1985) A mathematical model for the global
spread of influenza. Math Biosci 75: 3-22.

64. Hufnagel L, Brockmann D, Geisel T (2004) Forecast and control of
epidemics in a globalized world. Proc Natl Acad Sci USA 101: 15124-
15129.

65. Colizza V, Barrat A, Barth\`{e}lemy M, Vespignani A (2006) The role
of the airline transportation network in the prediction and predictability
of global epidemic. Proc Natl Acad Sci USA 103: 2015-2020.

66. Cooper BS, Pitman RJ, Edmunds WJ, Gay NJ (2006) Delaying the International
Spread of Pandemic Influenza. PLoS Med 3: e212.

67. Ovaskainen O, Cornell SJ (2006) Asymptotically exact analysis of stochastic
metapopulation dynamics with explicit spatial structure. Theor Popul Biol 69: 13-33.

68. Colizza V, Pastor-Satorras R, Vespignani A (2007) Reaction-diffusion
processes and metapopulation models in heterogeneous networks. Nat Phys 3:
276.

69. Epstein JM, Goedecke DM, Yu F, Morris RJ, Wagener DK, et al. (2007)
Controlling Pandemic Flu: The Value of International Air Travel Restrictions.
PLoS ONE 2: e401.

70. Colizza V, Barrat A, Barthelemy M, Valleron AJ, Vespignani A (2007) Modeling the
worldwide spread of pandemic influenza: Baseline case and containment interventions.
PLoS Med 4: e13.

71. Cornell SJ, Ovaskainen O (2008) Exact asymptotic analysis for metapopulation
dynamics on correlated dynamic landscapes. Theor Popul Biol 74: 209-225.

72. Balcan D, Colizza V, Gon\c{c}alves B, Hu H, Ramasco JJ, et al. (2009) Multiscale
mobility networks and the spatial spreading of infectious diseases. Proc Natl Acad
Sci USA 106: 21484-21489.

73. Vergu E, Busson H, Ezanno P (2010) Impact of the Infection Period Distribution
on the Epidemic Spread in a Metapopulation Model. PLoS ONE 5: e9371.

74. Balcan D, Vespignani A (2011) Phase transitions in contagion processes
mediated by recurrent mobility patterns. Nat Phys. doi:10.1038/nphys1944

75. United States Office of Management and Budget(OMB), OMB Bulletin No. 10-02:
Update of Statistical Area Definitions and Guidance on Their Uses(December 1, 2009).
Whitehouse website.\\
http://www.whitehouse.gov/sites/default/files/omb/assets/bulletins/b10-02.pdf. Accessed
2011 May 24.

76. Bureau of Transportation Statistics(BTS), United States, Air Carrier Traffic and Capacity
Data by On-Flight Market report(December 2009). BTS website. http://www.bts.gov/. Accessed
2011 May 24.

77. United States Census Bureau(CB), Annual Estimates of the Population of Metropolitan
and Micropolitan Statistical Areas: April 1, 2000 to July 1, 2009. U.S. Census Bureau website.\\
http://www.census.gov/popest/metro/. Accessed 2011 May 24.

78. United States Census Bureau(CB), American Factfinder. U.S. Census Bureau website.\\
http://factfinder.census.gov/home/saff/main.html?\_lang=en. Accessed 2011 May 24.

79. Centers for Disease Control and Prevention(CDC), United States, Swine Influenza A (H1N1)
Infection in Two Children --- Southern California, March-April 2009. CDC website.\\
http://www.cdc.gov/mmwr/preview/mmwrhtml/mm5815a5.htm. Accessed 2011 May 24.

80. Fraser C, Donnelly CA, Cauchemez S, Hanage WP, Kerkhove MDV, et al. (2009)
Pandemic Potential of a Strain of Influenza A (H1N1): Early Findings. Science 324:
1557-1561.

81. Balcan D, Hu H, Gon\c{c}alves B, Bajardi P, Poletto C, et al. (2009) Seasonal
transmission potential and activity peaks of the new influenza A(H1N1): a Monte
Carlo likelihood analysis based on human mobility. BMC Med. 7: 45.

82. Lessler J, Reich NG, Brookmeyer R, Perl TM, Nelson KE, et al. (2009)
Incubation periods of acute respiratory viral infections: a systematic review.
Lancet Infect. Dis 9: 291-300.

83. Yang Y, Sugimoto JD, Halloran ME, Basta NE, Chao DL, et al. (2009)
The Transmissibility and Control of Pandemic Influenza A (H1N1) Virus.
Science 326: 729-733.

84. Bo\"{e}lle PY, Bernillon P, Desenclos JC (2009) A preliminary estimation
of the reproduction ratio for new influenza A(H1N1) from the outbreak in
Mexico, March-April 2009. Euro Surveill 14: 19205.

85. Nishiura H, Castillo-Chavez C, Safan M, Chowell G (2009)
Transmission potential of the new influenza A(H1N1) virus and its
agespecificity in Japan. Euro Surveill 14: 19227.



86. Bureau of Transportation Statistics(BTS), United States (2010) ``Summary
2009 Traffic Data for U.S and Foreign Airlines: Total Passengers Down 5.3 Percent from 2008''.
BTS website. http://www.bts.gov/. Accessed 2011 May 24.

87. den Broeck WV, Gioannini C, Gon\c{c}alves B, Quaggiotto M, Colizza V, et al.
(2011) The GLEaMviz computational tool, a publicly available software to explore realistic epidemic
spreading scenarios at the global scale. BMC Infect Dis 11: 37.

88. Colizza V, Vespignani A (2008) Epidemic modeling in metapopulation systems with
heterogeneous coupling pattern: Theory and simulations. J Theor Biol 251: 450.

\newpage

\section*{Tables}

\begin{table}[!ht]
\caption{\bf{The empirical results of the parameters $\theta$ and $\eta$, and their
 relevance at the early time(the period between April 30th and June 1st, 2009),
using 2009 Pandemic A(H1N1) data collected by the WHO.}}
\begin{center}
\begin{tabular}{l c c c}
\hline
\bf \ \ \ \ Date    &\bf $ \theta$ &\bf $\eta$ &\bf $\theta\cdot \eta$  \\
\hline
                    April 30th   & \small 3.12      & \small 0.349 & \small 1.046\\
                    May\ \  1st  & \small 3.23      & \small 0.349 & \small 1.127\\
                    May\ \  2th  & \small 3.00      & \small 0.349 & \small 1.047\\
                    May\ \  3th  & \small 3.32      & \small 0.349 & \small 1.159\\
                    May\ \  4th  & \small 2.93      & \small 0.349 & \small 1.022\\
                    May\ \  5th  & \small 3.29      & \small 0.349 & \small 1.148\\
                    May\ \  6th  & \small 3.35      & \small 0.349 & \small 1.169\\
                    May\ \  7th  & \small 3.5       & \small 0.349 & \small 1.222\\
                    May\ \  8th  & \small 3.39      & \small 0.349 & \small 1.183\\
                    May\ \  9th  & \small 3.2       & \small 0.349 & \small 1.117\\
                    May\ \ 10th  & \small 3.16      & \small 0.349 & \small 1.103\\
                    May\ \ 11th  & \small 2.96      & \small 0.349 & \small 1.033\\
                    May\ \ 12th  & \small 3.06      & \small 0.349 & \small 1.068\\
                    May\ \ 13th  & \small 2.96      & \small 0.349 & \small 1.033\\
                    May\ \ 14th  & \small 3.00      & \small 0.349 & \small 1.047\\
                    May\ \ 15th  & \small 3.07      & \small 0.349 & \small 1.071\\
                    May\ \ 16th  & \small 3.07      & \small 0.349 & \small 1.071\\
                    May\ \ 17th  & \small 2.95      & \small 0.349 & \small 1.030\\
                    May\ \ 18th  & \small 2.93      & \small 0.349 & \small 1.023\\
                    May\ \ 19th  & \small 2.98      & \small 0.349 & \small 1.040\\
                    May\ \ 20th  & \small 2.97      & \small 0.349 & \small 1.037\\
                    May\ \ 21th  & \small 2.92      & \small 0.349 & \small 1.019\\
                    May\ \ 22th  & \small 2.82      & \small 0.349 & \small 0.984\\
                    May\ \ 23th  & \small 2.77      & \small 0.349 & \small 0.967\\
                    May\ \ 26th  & \small 2.62      & \small 0.349 & \small 0.914\\
                    May\ \ 27th  & \small 2.54      & \small 0.349 & \small 0.886\\
                    May\ \ 29th  & \small 2.44      & \small 0.349 & \small 0.852\\
                    June\  1st   & \small 2.33      & \small 0.349 & \small 0.813\\
\hline
\end{tabular}
\end{center}
\label{tab.1}
\end{table}

\begin{table}[!ht]
\caption{\textbf{The value of the parameters $\theta'_{us}$ and $\eta'_{us}$
for the simulation results at the early time of the period between $t=26$ and $t=39$.}}
\begin{center}
\begin{tabular}{l c c c}
\hline
\bf \emph{t}    &\bf $ \theta'_{us}$ &\bf $\eta'_{us}$ &\bf $\theta'_{us}\cdot \eta'_{us}$  \\
\hline
                    26   & \small 2.623     & \small 0.427 & \small 1.120\\
                    27   & \small 2.395     & \small 0.459 & \small 1.099\\
                    28   & \small 2.535     & \small 0.449 & \small 1.138\\
                    29   & \small 2.433     & \small 0.457 & \small 1.112\\
                    30   & \small 2.429     & \small 0.456 & \small 1.108\\
                    31   & \small 2.269     & \small 0.455 & \small 1.032\\
                    32   & \small 2.285     & \small 0.460 & \small 1.051\\
                    33   & \small 2.170     & \small 0.482 & \small 1.046\\
                    34   & \small 2.220     & \small 0.477 & \small 1.059\\
                    35   & \small 2.086     & \small 0.492 & \small 1.026\\
                    36   & \small 1.976     & \small 0.503 & \small 0.994\\
                    37   & \small 1.977     & \small 0.504 & \small 0.996\\
                    38   & \small 1.717     & \small 0.540 & \small 0.927\\
                    39   & \small 1.644     & \small 0.538 & \small 0.884\\

\hline
\end{tabular}
\end{center}
\label{tab.2}
\end{table}

\newpage

\newpage

\section*{Supporting Information}

{\Large
\textbf{Supporting Information Text S1}
}

\bigskip

{\Large\bf ``Evolution of scaling emergence in large-scale spatial epidemic spreading"}

\bigskip

{\bf Lin Wang, Xiang Li$\ast$, Yiqing Zhang, Yan Zhang, Kan Zhang}

{\bf $\ast$ Corresponding author: lix@fudan.edu.cn}

\subsection*{Simulation Model Design}

\ \ \ \ As a basic modeling scheme, we use the metapopulation approach, which explicitly
considers the geographical structure in the model by introducing multiple subpopulations
coupled by individuals' mobility. More specifically, the subpopulations correspond to the
metropolitan areas, and the dynamics of individuals' mobility is described by enplaning
between any two regions.

\ \ \ \ (\emph{i}) Infection Dynamics in a Single Subpopulation.

\ \ \ \ The infection dynamics takes place within each single subpopulation, and is described
by a homogeneously mixed population with an influenza-like illness compartmentalization
in which each individual exists in just one of the following discrete classes such as
susceptible(S), latent(L), infectious(I) or permanently recovered(R). In each
subpopulation $j$, the population is $N_{j}$, and $\mathcal {D}_{j}^{[\varphi]}(t)$ is the
number of individuals in the class $[\varphi]$ at time $t$. By definition, it is evident
that $N_{j}=\sum_{\varphi}\mathcal{D}^{[\varphi]}_{j}(t)$. Two essential kinds of the disease
evolution processes are considered in the infection dynamics: the contagion process(e.g.,
a susceptible individual acquires the infection from any given infectious individual and
becomes latent with the rate $\beta$, where $\beta$ is the transmission rate of a disease)
and the spontaneous transition of individual from one compartment to another(i.e. latent
ones become infectious with a probability $\epsilon$, or the infectious individuals recover
with a probability $\mu$, where $\epsilon^{-1}$ and $\mu^{-1}$ are the average latency time
and the average infection duration, respectively). Schematically, the stochastic infection
dynamics is given by
\begin{eqnarray}
(\mathcal {D}^{[S]}_{j},\mathcal {D}^{[L]}_{j},\mathcal
{D}^{[I]}_{j},\mathcal {D}^{[R]}_{j})\nonumber
\Rightarrow\left\{
\begin{array}{l@{} l}
\displaystyle (\mathcal {D}^{[S]}_{j}-1,\mathcal
{D}^{[L]}_{j}+1,\mathcal {D}^{[I]}_{j}, \mathcal {D}^{[R]}_{j}),&\
with\  rate\  \beta \mathcal {D}^{[S]}_{j}\mathcal
{D}^{[I]}_{j}/N_{j},\\[0.1cm]
\displaystyle(\mathcal {D}^{[S]}_{j},\mathcal {D}^{[L]}_{j}-1,
\mathcal {D}^{[I]}_{j}+1,\mathcal {D}^{[R]}_{j}),&\ with\  rate\
\epsilon \mathcal {D}^{[L]}_{j},\\[0.1cm]
\displaystyle (\mathcal {D}^{[S]}_{j},\mathcal
{D}^{[L]}_{j},\mathcal {D}^{[I]}_{j}-1, \mathcal {D}^{[R]}_{j}+1),&\
with\  rate\  \mu \mathcal {D}^{[I]}_{j}, \nonumber
\\
                \end{array}
\right.
\end{eqnarray}
where the first reaction reflects the fact that each susceptible in subpopulation $j$
would be infected by contacting any infectious individuals with probability
$\beta\mathcal {D}^{[I]}_{j}/N_{j}$, therefore the number of new infections generated
in subpopulation $j$ at time $t+1$ is extracted from a binomial distribution with the
probability $\beta\mathcal {D}^{[I]}_{j}(t)/N_{j}(t)$ and the number of trials
$\mathcal {D}^{[S]}_{j}(t)$; the second and the third reactions represent the
spontaneous transition process.

\ \ \ \ (\emph{ii})Epidemic Transmission among Different Subpopulations.

\ \ \ \ As individuals travel around the country, the disease may spread from one area to
another. Therefore, in addition to the infection dynamics taking place inside each subpopulation,
the epidemic spreading at a large geographical scale is inevitably governed by the human
mobility among different subpopulations by means of the domestic air transportation. Since
the \emph{BTS} report reflects the actual aviation flows between any two U.S. airports,
we define a stochastic dispersal operator $\nabla_{j}(\{\mathcal {D}^{[\varphi]}\})$
representing the net balance of individuals in a given compartment $\mathcal {D}^{[\varphi]}$
that entered in and left from each subpopulation $j$. In each subpopulation $j$, the dispersal
operator is expressed as
\begin{eqnarray}
\nabla_{j}(\{\mathcal{D}^{[\varphi]}\})=\sum\limits_{\ell}(\mathcal {X}_{\ell j} (\mathcal {D}^{[\varphi]}_{\ell})-\mathcal {X}_{j\ell}(\mathcal {D}^{[\varphi]}_{j})),\nonumber
\end{eqnarray}
where $\mathcal {X}_{j\ell}(\mathcal {D}^{[\varphi]}_{j})$ describes the daily number of
individuals in the compartment $[\varphi]$ traveling from
subpopulation $j$ to subpopulation $\ell$. In the scenario of air travel, this operator is
relative to the passenger traffic flows and the population. Neglecting multiple legs travels
and assuming the well mixing of population inside each subpopulation, we deduce that the
probability of any individual traveling on each connection $j\rightarrow\ell$ everyday is
given by $p_{j\ell}=\omega_{j\ell}/N_{j}$, where $\omega_{j\ell}$ represents the
daily passenger number from $j$ to $\ell$. The stochastic variables $\mathcal {X}_{j\ell}
(\mathcal {D}^{[\varphi]}_{j})$ therefore follow the multinomial distribution
\begin{eqnarray}
P(\{\mathcal {X}_{j\ell}\})&=&\frac{\mathcal {D}^{[\varphi]}_{j}!}{(\mathcal
{D}^{[\varphi]}_{j}-\sum_{\ell}\mathcal {X}_{j\ell})!\prod_{\ell}\mathcal
{X}_{j\ell}!}(\prod \limits_{\ell}p_{j\ell}^{\mathcal {X}_{j\ell}}) (1-\sum\limits_{\ell}p_{j\ell})^{(\mathcal {D}^{[\varphi]}_{j}-\sum_{\ell}\mathcal {X}_{j\ell})},\nonumber
\end{eqnarray}
where $(\mathcal {D}^{[\varphi]}_{j}-\sum_{\ell}\mathcal {X}_{j\ell})$ indicates daily number
of non-traveling individuals of compartment $[\varphi]$ staying in subpopulation $j$. It is
noticeable that the population $N_{j}$ of each subpopulation keeps constant, e.g.,
$\sum_{[\varphi]}\nabla_{j}(\mathcal {D}^{[\varphi]})=0$, because the passenger flows are
balanced on each pair of connections in this article.

\section*{Pandemic phases defined by the WHO}

\ \ \ \ 1. Interpandemic period

\ \ \ \ Phase 1: no new influenza virus subtypes circulating among animals have been reported to cause
infections in humans.

\ \ \ \ Phase 2: a new influenza virus subtypes circulating among domesticated or wild animals is known
to have caused infection in humans, and is therefore considered a potential pandemic threat.

\ \ \ \ 2. Pandemic alert period

\ \ \ \ Phase 3: a new influenza virus subtypes has caused sporadic cases or small clusters of disease
in people, but has not resulted in human-to-human transmission sufficient to sustain
community-level outbreaks.

\ \ \ \ Phase 4: it is characterized by verified human-to-human transmission of a new influenza virus
subtypes able to cause ``community-level outbreaks". Though the virus is not well adapted to
humans, the ability to cause sustained disease outbreaks in a community marks a significant
upwards shift in the risk for a pandemic.

\ \ \ \ Phase 5: it is characterized by human-to-human spread of the virus into at least two countries
in one WHO region. While most countries will not be affected at this stage, the declaration of
Phase 5 is a strong signal that a pandemic is imminent.

\ \ \ \ 3. Pandemic period

\ \ \ \ Phase 6: it is characterized by community level outbreaks in at least one other country in a
different WHO region in addition to the criteria defined in Phase 5. This phase indicates that
a global pandemic is under way.

\section*{Power Law Distribution with an Exponential Cutoff}

\ \ \ \ In fact, little real systems do display a perfect power law pattern for the
Zipf's distribution or probability density distribution[1,2]. When an exponential cutoff
is added to the power law function, the fit is substantially improved for dozens of
systems, e.g., the forest fires, earthquakes, web hits, email networks, sexual contact
networks. The cutoff indicates that there is a characteristic scale, and that infrequently
super-enormous events are exponentially rare. Strictly speaking, a cutoff power law should
always fit the data at least as good as a pure one(just let the cutoff scale go to infinity),
thus the power law distribution can be deemed as a subset of the exponentially cutoff power law[2].

\section*{The Mass Action Principle}

\ \ \ \ Prof. Hamer postulated that the course of an epidemic depends on the rate of
contact between susceptible and infectious individuals. This conception plays
a significant role in the study of epidemiology; it is the so-called `mass
action principle' in which the net rate of spread of infection is assumed to
be proportional to the product of the density of susceptible persons times the
density of infectious individuals[3,4].

\bigskip

\bigskip

1. Newman MEJ (2005) Power laws, Pareto distributions and Zipf's law.
Contemporary Physics 46: 323-351.

2. Clauset A, Shalizi CR, and Newman MEJ (2009) Power-law distributions
in empirical data. SIAM Review 51, 661-703.

3. Anderson RM and May RM (1991) Infectious Diseases of Humans: Dynamics
and Control(Oxford Unvi. Press, Oxford).

4. Hamer WH (1906) The Milroy Lectures On Epidemic disease in England ---
The evidence of variability and of persistency of type. The Lancet 167:
733-739.

\section*{Figure Legends}

\setcounter{figure}{0}

\begin{figure}[!ht]
\makeatletter
\renewcommand{\thefigure}{S\@arabic\c@figure}
\begin{center}
\includegraphics[width=6in]{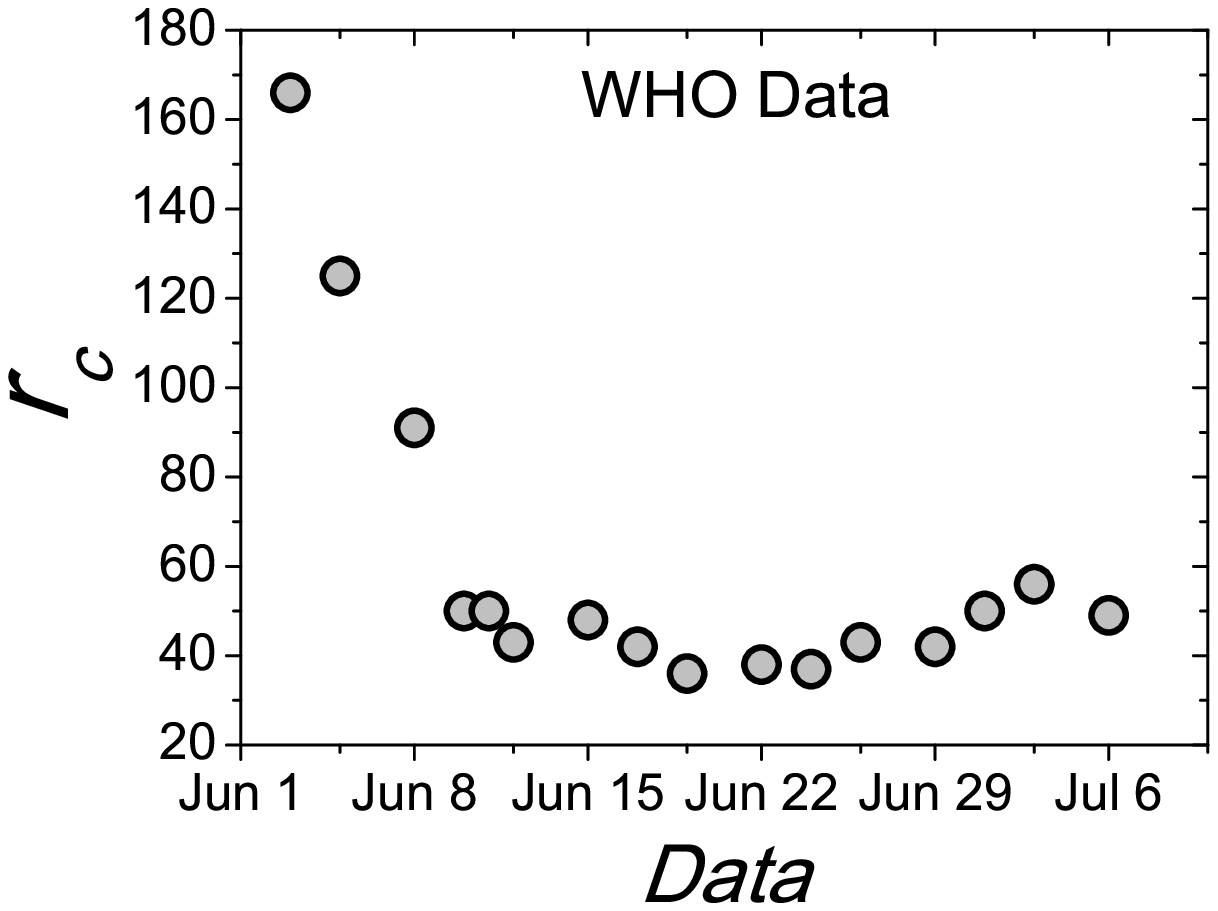}
\end{center}
\caption{\bf The temporal evolution of the estimated parameter $r_{c}$, data provided by the WHO.}
\label{fig.S1}
\end{figure}

\begin{figure}[!ht]
\makeatletter
\renewcommand{\thefigure}{S\@arabic\c@figure}
\begin{center}
\includegraphics[width=6in]{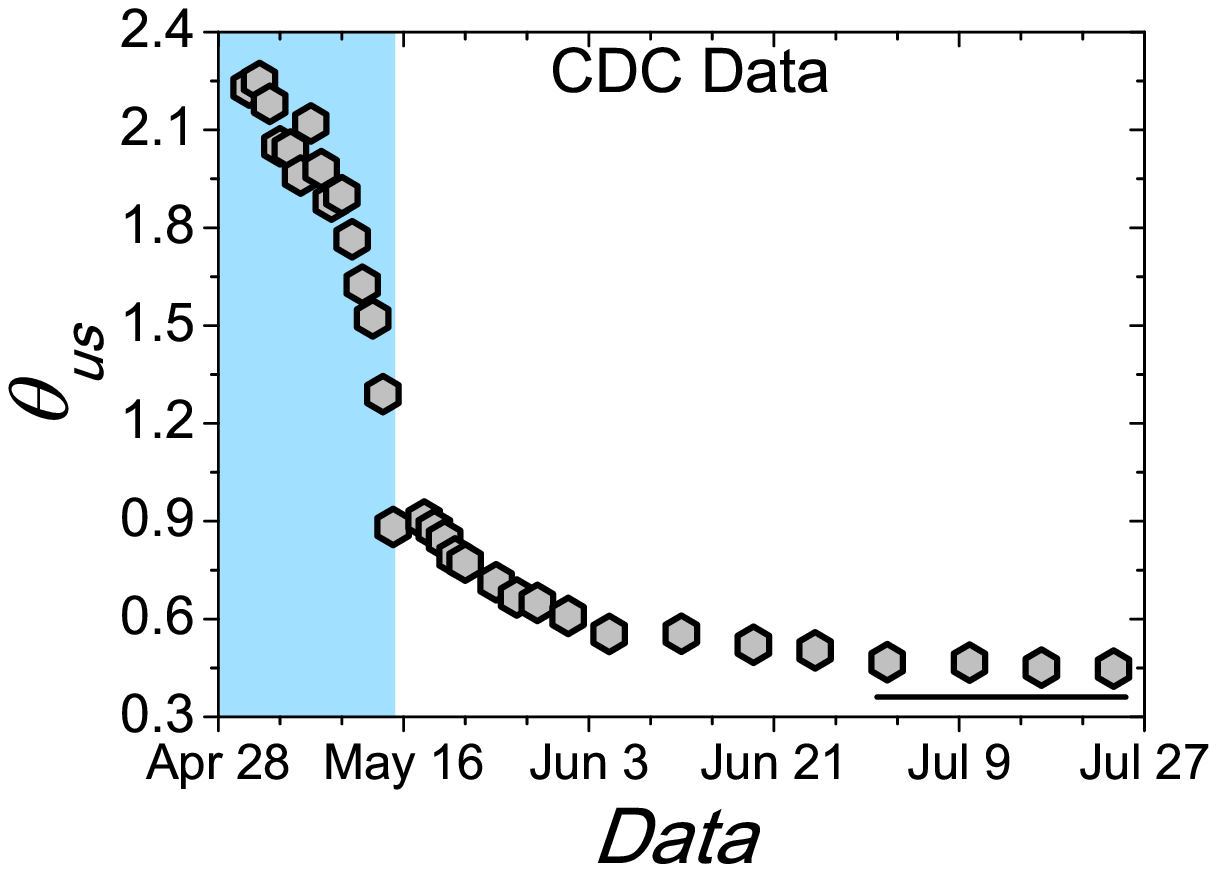}
\end{center}
\caption{\bf The temporal evolution of the estimated exponent $\theta_{us}$ for all data provided
by the CDC.}
\label{fig.S2}
\end{figure}

\begin{figure}[!ht]
\makeatletter
\renewcommand{\thefigure}{S\@arabic\c@figure}
\begin{center}
\includegraphics[width=5in]{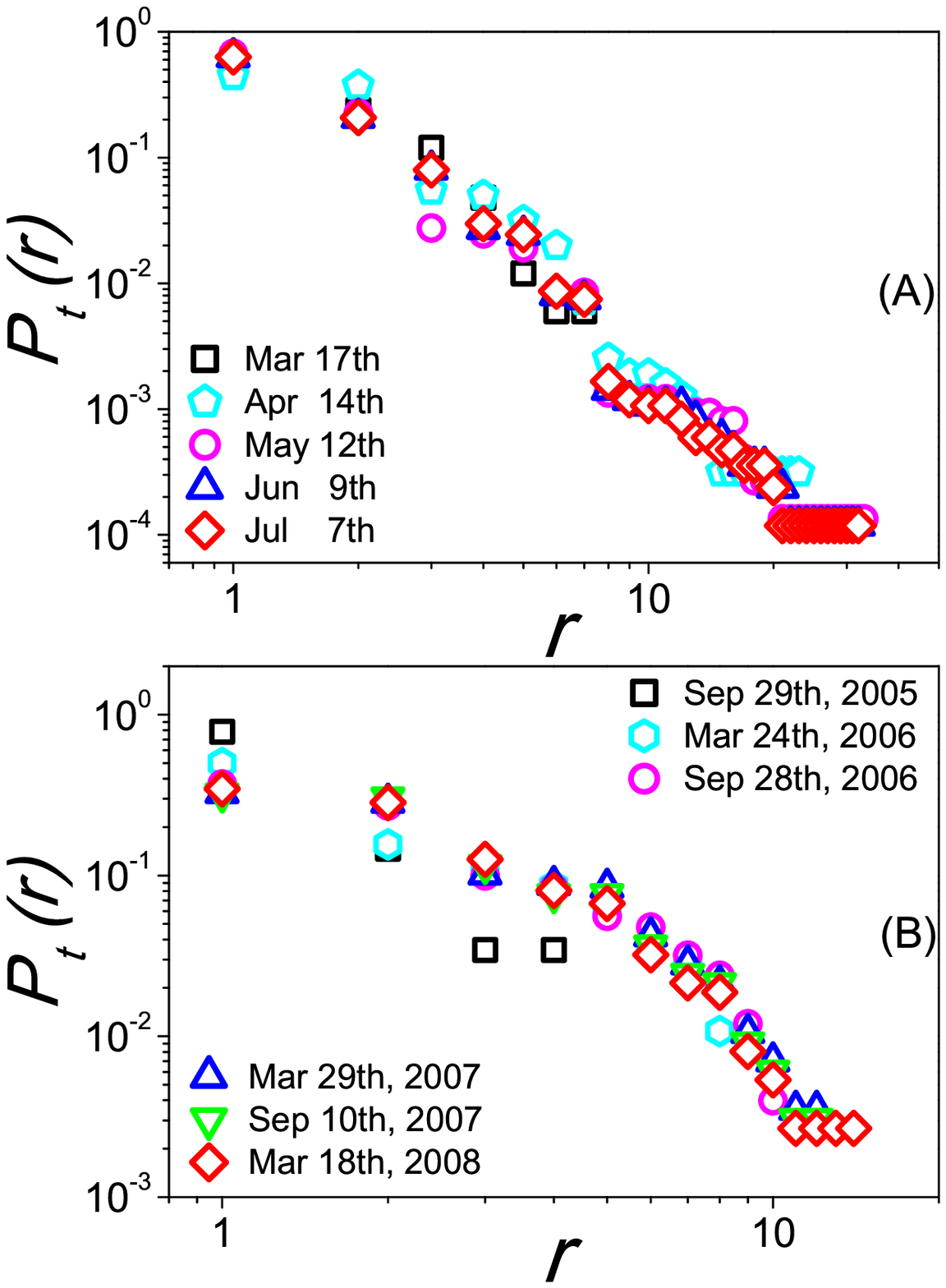}
\end{center}
\caption{\bf The empirical results of the SARS and avian influenza(H5N1).
(A) shows the normalized probability-rank distribution of the cumulated confirmed number
of every infected country around the world at several given date sampled about every four
weeks, data provided by the WHO(http://www.who.int/csr/sars/country/en/index.html).
(B) shows the normalized probability-rank distribution of the cumulated confirmed number
of every infected country around the world at several given date sampled about every half
a year, data provided by the WHO(http://www.who.int/csr/disease/avian\_influenza\\/country/en/).}
\label{fig.S3}
\end{figure}

\end{document}